\documentclass[preprint2]{aastex}
\lefthead{THORSTENSEN ET AL.}
\righthead{}
\begin{document}
\title{The Dwarf Novae of Shortest Period
\footnote{Based in part on
observations obtained at the Michigan-Dartmouth-MIT Observatory.}
}

\author{John R. Thorstensen}
\affil{Department of Physics and Astronomy\\
6127 Wilder Laboratory, Dartmouth College\\
Hanover, NH 03755-3528\\
electronic mail: {\tt thorstensen@dartmouth.edu}}
\author{Joseph O. Patterson, Jonathan Kemp}
\affil{Department of Astronomy, Columbia University\\
538 West 120th Street, New York, NY 10027\\
electronic mail: {\tt jop@astro.columbia.edu}}
\author{St\'ephane Vennes}
\affil{Astrophysical Theory Centre\\
Australian National University, ACT0200, Australia\\
electronic mail: {\tt vennes@wintermute.anu.edu.au}}

\begin{abstract}
We present observations of the dwarf novae GW Lib, V844 Her,
and DI UMa.  Radial velocities of H$\alpha$ yield
orbital periods of $0.05332  \pm 0.00002$ d (= 76.78 m) for GW Lib and
and $0.054643 \pm 0.000007$ d (= 78.69 m) for V844 Her.
Recently, the orbital period of DI UMa
was found to be only $0.054564 \pm 0.000002$ d (= 78.57 m)
by \citet{fr99}, so these are the 
three shortest orbital periods among dwarf novae
with normal-abundance secondaries.

GW Lib has attracted attention as a cataclysmic binary
showing apparent ZZ Ceti-type pulsations of the 
white dwarf primary.  Its spectrum shows 
sharp Balmer emission flanked by strong, broad
Balmer absorption, indicating a dominant
contribution by white-dwarf light.  
Analysis of the Balmer absorption
profiles is complicated by the unknown residual
accretion luminosity and lack of coverage of the 
high Balmer lines.  Our best-fit model atmospheres 
are marginally hotter than the ZZ Ceti instability strip, in 
rough agreement with recent ultraviolet results from
HST.  The spectrum and outburst 
behavior of GW Lib make it a
near twin of WZ Sge, and we estimate it to have a 
quiescent $M_V \sim 12$.  Comparison with archival
data reveals proper motion of $65 \pm 12$ mas yr$^{-1}$. 

The mean spectrum of V844 Her is typical of SU UMa
dwarf novae.  We detected
superhumps in the 1997 May superoutburst
with $P_{\rm sh} = 0.05597 \pm 0.00005$ d.
The spectrum of DI UMa appears normal for a dwarf nova
near minimum light.

These three dwarf novae have nearly identical short periods
but completely dissimilar outburst characteristics. 
We discuss possible implications.

\end{abstract}

\section{Introduction}

Cataclysmic variable stars are close binary systems
in which mass is transferred
from a low-mass secondary onto a white dwarf;
\citet{warn} has written an 
excellent monograph on cataclysmics.
In the dwarf nova subclass, the transferred
matter is thought to accumulate in an accretion disk until a
critical density is reached, whereupon the transfer of matter
through the disk dramatically increases, causing a brightening
of the star.  The outbursts of dwarf novae range widely in amplitude
and frequency, with the less-frequently outbursting systems tending to 
have the greatest amplitudes (the Kukarkin-Parenago relation). 

The orbital periods $P_{\rm orb}$ of dwarf novae range from $\sim 12$ hours
down to about 75 minutes, near the theoretical minimum period
for a hydrogen-rich donor star \citep{rjw82}.
(We do not here consider systems with helium
donor stars, which can reach much shorter periods.)  Dwarf novae with 
$P_{\rm orb}$ less than about 3 h (effectively, those shortward
of the 2-3 h period `gap') show occasional large-amplitude, long-duration
outbursts, called superoutbursts,
which are accompanied by photometric oscillations, called 
superhumps, which have periods $P_{\rm sh}$ a few percent {\it longer}
than $P_{\rm orb}$.  Superoutbursters are
classified as SU UMa stars, after their prototype.
But SU UMa stars
show a wide range of intervals between outbursts, from
about 4 days for ER UMa and its brethren \citep{jc,p95,rht95,ms95,twoers},
to several decades
for WZ Sge and similar stars 
\citep{wxcet,alcom,egcnc,katoeruma95}.   

Here we report extensive observations of
the dwarf novae GW Lib and V844 Her, 
which prove to have very short orbital periods.  
We also present a spectrum of DI UMa, another very
short period system.  In spite of their similar 
orbital periods, these three systems 
have quite different outburst characteristics.  

Section 2 of this paper describes observational protocols
and the analysis which produced orbital periods for 
GW Lib and V844 Her.  Sections 3, 4, and 5 contain more detailed
discussions of GW Lib, V844 Her, and DI UMa respectively.   
Finally, Section 6 contrasts the three systems and discusses 
implications for cataclsymic binary evolution. 

\section{Observations}

Our spectroscopy is from the 2.4 m Hiltner telescope 
at MDM Observatory.  
The modular spectrograph and a $2048^2$ CCD detector
gave spectra covering 4300 -- 7500 \AA\ at 2 \AA\
pixel$^{-1}$, and a 1 arcsec slit gave 3.5 \AA\ resolution
over most of the range.
We observed flux standards when appropriate, took comparison lamps
frequently as the telescope tracked, and reduced the
data using standard procedures in IRAF.  A few spectra were
adjusted slightly in velocity when checks of the
$\lambda 5577$ night-sky line revealed that a comparison 
lamp exposure had gone awry.  From these checks
we estimate our velocity scale to be correct to within 
$\sim 10$ km s$^{-1}$ and rather more stable than this.
The flux scale suffers from occasional clouds, uncalibrated
losses at the spectrograph slit, and a poorly-understood
instrumental problem which causes inconsistent continuum
shapes among the standard stars.  The spectra shown here
are averages of many, which appears to have ameliorated these
problems.  Figs.~1 and 2 show averaged
spectra of all the objects,
and Table 1 gives quantitative measures of the spectra. 

For GW Lib and V844 Her, our spectra were extensive enough for
us to derive orbital periods.
We measured velocities of the H$\alpha$ emission using convolution
techniques \citep{sy}, and searched for periods
among the resulting time series (Fig.~3) using the sinusoidal fit method
described by \citet{tpst}.  To asses the reliability
of alias choices we used the Monte Carlo test of \citet{tf}. 
Table 2 (available in full in the electronic
version of this paper) lists the radial velocities, 
Table 3 gives sinusoidal fits to the velocities at the
preferred periods of the form 
$$v(t) = \gamma + K \sin[2 \pi (t - T_0) / P],$$
and Fig.~4 shows the velocities folded at the best
periods, with the fits superposed.
In Table 3, $\sigma$ is the uncertainty of a single velocity
derived from the goodness of fit.  Because the emission
lines may not reflect the motion of either star accurately,
we caution against using $\gamma$ or $K$ for dynamics.
However, experience suggests that the spectroscopic 
$P$ should be a reliable measure of $P_{\rm orb}$. 

Our photometric data are more various.
We obtained time-series photometry of the outburst of V844 Her
using the Center for Backyard Astrophysics
network of photometrists (e.g., \citealt{egcnc}).
In addition, on 2000 Jan.\ 11 UT we obtained $UBVRI$
images of GW Lib with the Hiltner 2.4 m  
and the central 1024$^2$ pixels of 
a SITe $2048^2$ CCD, at a scale of 0.275 arcsec pixel$^{-1}$. 
These were calibrated with observations of 
\citet{landolt} standard stars.

\section{GW Lib}

GW Lib was discovered in 1983 as a 9th magnitude novalike 
object \citep{gm83}.  After the object faded to $V \sim 16$, 
\citet{ds87} found a strong blue continuum with relatively 
sharp Balmer emission lines.  \citet{duerbeck}
lists its classification as uncertain between nova
or dwarf nova.  \citet{rnova96} also show a spectrum,
note its similarity to WZ Sge, and classify GW Lib as a
dwarf nova.  The outburst recurrence time is likely to be very 
long ($\ge 10$ yr), though outbursts could have gone undetected 
given the star's low ecliptic latitude ($-6^{\circ}.4$).  

\citet{vanzyl02} report extensive time-series photometry of 
GW Lib in 1997, 1998, and 2001, and found dominant oscillations
at periods near 650, 370, and 230 s.  The pulsation spectrum
is unstable, with many signals closely spaced in frequency.
They interpret these signals as arising from non-radial
pulsations of the white dwarf, making it the only such pulsator
also known to be accreting.  
A report of their early results prompted us to observe the star
in 1999 June.

\citet{szkodygw} report spectroscopy of GW Lib,
from which they derived an orbital period of $79.4 \pm 0.3$ 
minutes and found the signature of a white dwarf photosphere
in the mean spectrum.  Our results described below are in
broad agreement, but we do find a significantly shorter
orbital period of $76.79 \pm 0.02$ min.  Many of our observations
occurred on adjacent nights, allowing cycles to be counted from
night to night, but their three nights were separated too 
widely in time to allow this.  Thus the present
determination supersedes that of \citet{szkodygw}.   
\citet{szkody02} also obtained an HST spectrum in 2002 January
which indicates a white-dwarf temperature of 14,700 kelvin; 
our less precise measure (see below) is in satisfactory agreement
with this determination.

Intriguingly, \citet{woudt02} detected
a 2.09-h photometric modulation in observations obtained 
2001 May (but not in observations obtained at other times). 
They note that this periodicity has no particular relationship
to the orbital period, and that its cause is unknown.

\subsection{Spectroscopy}

Because of GW Lib's southerly declination, we pushed the observations to 
airmasses as large as 4.6 to obtain the largest possible
range of hour angle, and hence the greatest discrimination
between daily cycle-count aliases.  The resulting data 
set spans 5.5 h, and includes 88 exposures of 300 s each.
In the mean spectrum (Fig.~1) 
the continuum at $\lambda 5500$ corresponds to 
$V \sim 17.4$; because an unknown fraction of the
light is lost at the slit, this 
is in fair agreement with the magnitudes derived from the 
CCD photometry, namely 
$V = 17.01, B - V = 0.12, U - B = -0.50,$ and $V - I_{KC} = 0.12$.
The strong blue continuum and Balmer absorption
wings suggest that the bulk of the light arises in a 
white dwarf photosphere.  The Balmer emission lines are 
unusually narrow, with Gaussian fits giving $\sim 10$ \AA\ 
FWHM (full width half maximum).  Weak HeI emission is
visible at $\lambda \lambda$ 5876, 6678, and 7027.  
An emission feature near $\lambda$ 5169 is present, which
is usually attributed to FeII, and the Na D lines appear
weakly in emission.  All other features are artifacts; they 
include incompletely subtracted night sky lines at $\lambda \lambda$ 5577
and 6300, and a telluric absorption band at $\lambda$ 6870
which did not divide out completely.

The convolution function used for the velocities was optimized
for 10 \AA\ FWHM, so our velocities pertain to the narrow line core.  
The periodogram (Fig.~3) indicates an orbital 
frequency near 18.75 cycles d$^{-1}$, and the Monte Carlo test 
shows that the choice of daily cycle
count is secure at $> 99$ per cent confidence.
The orbital period -- $76.79 \pm 0.02$ min -- 
is the shortest known for dwarf nova accreting from companions
with normal hydrogen abundances
(the secondary stars in shorter-period objects such as 
RX2329+06 are probably enriched in helium;
\citealt{thor02}).

\subsection{Binary Inclination, Accretion Rate, Distance, and Proper Motion}

The long outburst interval, short orbital period, and spectral
appearance of GW Lib make it an excellent match for WZ Sge.  In
Fig.~1 we present a spectrum of WZ Sge for comparison --
they appear very similar, except the emission lines of GW Lib
are much narrower.  

The narrowness of the emission lines suggests a low binary inclination.
WZ Sge has $i \sim 77^{\circ}$ 
\citep{spruitwzsge}, and the emission lines 
in our spectrum of WZ Sge are about five times wider
than those in GW Lib.  Attributing the difference entirely
to inclination leads to an estimate of $i \approx 11^{\circ}$.
The emission in WZ Sge has a strong S-wave component;
we do not resolve this component in GW Lib (indeed, we
don't see any clear substructure in the emission lines), 
but an S-wave contribution may be 
dominating the apparent motion of the line.  

Apparently, no accurate photometry was obtained during GW Lib's 
only known outburst, 
but the photographic magnitude at discovery was estimated
as 9.0.  Since $B - V \approx 0.$ for dwarf novae in outburst, 
we take the outburst $V$ magnitude to be near 9.  
\citet{warn} presents a standard-candle relationship
for dwarf novae in outburst, which predicts an
$M_V = +5.4$ at this $P_{\rm orb}$.
Superoutbursts tend to be about
1 mag brighter than this, so we estimate 
$M_V = +4.5$.  Warner's relation is reckoned at $i = 56^{\circ}$, but
GW Lib is evidently more face-on than this, 
making it about 1 magnitude brighter than at the canonical
inclination, or $M_V = +3.5$.  This would put GW Lib
at only $\sim 125$ pc distance.  Our quiescent
$V = 17.0$ then implies $M_V \approx +11.5$. 

Because of the short distance indicated by the above arguments,
we searched for a proper motion.  The Digitized Sky Survey has
two epochs available, 1976.3 and 1991.3.  GW Lib was also imaged in
the original Palomar Observatory Sky Survey (epoch 1955.4 for this field),
and Dr. David Monet of USNO kindly provided us with coordinates
from this survey measured with the USNO Plate Measuring Machine.
We use our I-band 2.4m CCD image for the final epoch (2000.02).  
The CCD image gave the best internal precision, so we fit centroids
from this image to the USNO A2.0 catalog to derive refined 
nominal standard coordinates for 99 field stars, and then 
transformed coordinates from the other images onto this system
using 6-constant plate model fits.  GW Lib showed a significant
proper motion $\mu = 66 \pm 12$ mas yr$^{-1}$ in position angle   
$296 \pm 10$ degrees.  This corroborates a short distance; its
transverse velocity would be 39 km s$^{-1}$ at 110 pc.
The celestial coordinates of GW Lib for epoch and equinox 2000 
refererred to the USNO A2.0 system (nominally ICRS) are
$\alpha = 15^{\rm h}\ 19^{\rm m}\ 55^{\rm s}.33$ and 
$\delta = -25^{\circ}\ 00'\ 24''.7$.

\subsection{Line Profile Analysis}

We estimated the white dwarf's atmospheric parameters
by fitting the Balmer line profiles to a grid of synthetic white-dwarf
spectra spanning $T_{\rm eff}$ from 10000 to 84000 kelvin and
$\log g$ from 7.0 to 9.0, with $\log ({\rm He/H})= -9$.  
These were computed using a new version of the 
\citet{mih75}  computer code modified to
include convection within the existing numerical treatment.
Potential convective instability of each depth layer was tested
against the standard Schwarzschild
criterion and the convective flux was computed using the 
mixing length formalism \citep{mih78}. We adopted convective
parameters appropriate for ZZ Ceti white dwarfs following the analysis of
\citet{ber95}, namely the ML2 parameters with the mixing length
over pressure scale height ratio $\alpha$ adjusted to 0.6.
Neutral hydrogen b-b, b-f, and f-f opacities
as well as b-f and f-f opacities of the negative hydrogen ion are included
in the calculation. The model atmospheres are converged until the total
flux is constant to at least one part in $10^6$. 
Detailed Stark-broadened synthetic line profiles are computed 
based on the converged models.

Fig.~5 shows the best fit to the H$\alpha$ and H$\beta$ line profiles. The
central emission is excluded from the fit ($\pm 50 \AA$).
A general search between $T_{eff}=10,000$ K and 30,000 K reveals
a best solution at $T_{eff}=24,000$ and log $g=8.76$. This solution is
totally incompatible with the optical spectral energy distribution which
imposes a much cooler temperature. Restricting the fit to the range of
temperature between 10,000 and 14,000 K reveals a solution close to 13,000 K.
A slightly different solution is obtained when excluding $\pm$40\AA\
from the H$\alpha$ core, and $\pm$30 \AA\ from H$\beta$.
A comparison of the two solutions underlines systematic effects due to the
incompleteness of the data set.  Adding possible continuum disk 
contamination to the synthetic spectra in the form of a
contribution corresponding to 5\% and above of the observed flux at 5500\AA,
eliminates all solutions below 16,000 kelvin, so the disk contamination
appears to be small.

The best-fit parameters ($T_{eff}=13220$K, log$g=7.4$) are above
the ZZ Ceti instability strip but the 90\% confidence contour extends
within the range ascribed to the pulsating class of DA white dwarfs by 
\citet{ber95}, using the ML2 $\alpha=0.6$ formalism. 
The results of the present
analysis are uncertain, because 
(1) systematic effects such as disk contamination
are not well ascertained and (2) the quality of the line profile fit is
not satisfactory and is based on only two lines.  As noted earlier,
\citet{szkody02} find a temperature of 14,700 K for the white dwarf based
on an HST UV spectrum; our result is consistent with theirs, but
is likely to be less accurate for the reasons noted above.

\section{V844 Herculis}

     V844 Herculis is a dwarf nova discovered by 
\citet{a96} in a
search for variable stars in a photographic survey with the 40 cm
Crimean astrograph.  On 419 plates in the interval 1960-1994, Antipin
found variations in the range B=12.5-17.5, with the best-observed
outburst lasting 12-18 days.  This announcement captured the attention
of variable-star observers, and the star has now become frequently
monitored by visual observers.  Long outbursts were seen in May 1997
and December 1998, and the outburst of October 1996 was probably also
long.  The interval between these long outbursts is not securely known,
but is probably about 280 days.

\subsection{The May 1997 Outburst}

     News of the May 1997 outburst drove the CBA
network into action.  Starting with the third night of outburst, we
obtained photometric coverage on 21 of 37 nights, following the star
all the way down to quiescence.  The photometry log is given in Table
4.

     The eruption timescale (15 d bright, followed by rapid return to
quiescence in 1-2 d) is typical of superoutbursts in SU UMa stars.
Even more characteristic are the photometric waves (superhumps)
prominent throughout
the eruption.  Fig.~6 (upper panel) shows the superhumps in one long night of 
coverage.  These establish the star's
membership in the SU UMa class of dwarf novae.

     We calculated the mean superhump amplitude and time of maximum
light for each night of observation.  These measurements are given in
the final columns of Table 4.  The amplitude spectrum of the 10-day
superoutburst light curve (JD 2450592-602) is shown in the 
lower panel of Fig.~6, indicating
a powerful signal at $17.87 \pm 0.02$ cycle d$^{-1}$ and its first harmonic.  
A signal is seen near 71.88 cycle d$^{-1}$ also.  Since we know the 
orbital frequency to be
$\omega = 18.300$ cycle d$^{-1}$ (see below), 
we interpret 17.87 cycle d$^{-1}$ as 
the basic superhump frequency,
the $\omega - \Omega$ lower precessional sideband of the 
orbital frequency.  Then the
35.75 cycle d$^{-1}$ signal is the first harmonic, at $2\omega - 2 \Omega$.
The identification of
the 71.88 cycle d$^{-1}$ signal is less clear, but it could 
be the $4\omega - 3 \Omega$ component
(in general, superhumps come with $n \omega - m \Omega$ 
components, where $m$ and $n$ are any
small integers and $m \le n$; Skillman et al. 1999).  
The superhump {\it changes} 
measurably in frequency and waveform during the 10-day interval, so precise
identification of higher components is somewhat ambiguous.

\subsection{Spectroscopy} 

Our spectra are from 1997 June and July, after the 
star had returned to quiescence.  The observations, comprising
158 300-s exposures, span 18 d of
elapsed time and 7.06 h of hour angle.
The mean spectrum (Fig.~2) is normal for a dwarf nova; in 
particular, the hydrogen and helium lines appear similar to 
other dwarf novae of longer period, emphasizing that the 
accreted material must come from the usual hydrogen-envelope
secondary despite the short orbital period.
The emission lines are double-peaked; in H$\alpha$ 
and H$\beta$, the peaks are
separated by $\pm 400$ km s$^{-1}$ and at the half-intensity
points the lines extend to $\pm 900$ km s$^{-1}$.
The continuum level implies $V \sim 17.9$, with an
uncertainty of a few tenths from unknown light
losses at the spectrograph slit.
The H$\alpha$ emission velocities (Table 2) were measured using 
the derivative of a Gaussian as the convolution function; 
this was optimized for a 40 \AA\ line width, a little wider than the
36 \AA\ observed width of H$\alpha$, in order to emphasize
the wings of the line profile.  We found that the velocity amplitude
was strongly affected by the algorithm used,
reinforcing the usual cautions about using CV emission lines
for dynamical mass estimates.  However, the period (Table 3) is 
determined without ambiguity.

\section{DI Ursae Majoris}

DI UMa \citep{fr99,kato96} has superoutbursts recurring on a
timescale of 30 to 45 d.  This places it among the most frequently
superoutbursting SU UMa stars, a group loosely referred to as the
ER UMa stars.
We did not accumulate sufficient spectra to find
an independent orbital period; photometry at quiescence 
\citet{fr99} yields a modulation at 
$0.054564 \pm 0.000001$ d (= 78.57 min), which is presumably the orbital
period.  Fig.~2 shows the mean spectrum,
and Table 1 lists the spectral features.  
The normality of the spectrum suggests that the abundances in
the accreting material are normal, which in turn suggests that
the secondary is not enriched in helium.

\section{Discussion}

     In Table 5 we list the seven
dwarf novae (or dwarf-nova candidates)
with the shortest well-established orbital periods.
Two of the systems have periods shorter than 
GW Lib.  V485 Cen has a dramatically 
short 59-min period.  \citet{aug96} explain this
by invoking a low hydrogen abundance for the secondary,
so it may not belong among the more usual hydrogen accretors.
Support for this idea comes from the apparently related
system RX2329+06, which has a K-type secondary at 
$P_{\rm orb} = 64$ min, a period at which
a CV secondary with normal hydrogen abundance would be
be much cooler.  \citet{thor02} show that
a helium-enriched secondary matches the observations,
and propose that mass transfer began near the end of
core hydrogen burning.
Another system, RX2353$-$38, resembles a dwarf nova 
spectroscopically, and it probably is one, but it has 
never been observed to erupt.  Thus the stars studied here 
have the three shortest periods among well-established, 
hydrogen-accreting dwarf novae.

In the most straightforward scenarios for CV 
evolution (see \cite{king88}
for a review),  the properties of dwarf novae are
regulated by the accretion rate, which is dependent on the mass of
the secondary, which depends on orbital period.   
Thus the outburst properties should depend in a simple way 
on $P_{\rm orb}$.  In particular,
stars of the shortest orbital period should have very rare outbursts,
like the well-studied prototype WZ Sge.

     The truth appears to be much more complex.  One of these stars, GW
Lib, is indeed an excellent match to WZ Sge: intrinsically very faint,
with a recurrence period probably exceeding 10 years.  Yet DI UMa is a rather
bright star (at the distance estimated by \cite{fr99}
it would have $M_V = +8.4$ at minimum), and one of the most frantic 
dwarf novae in the sky,
erupting at least 100 times more frequently.  And V844 Her is an
intermediate case, but unusual in never having shown any normal
outbursts (despite the modest interval of $\sim 280$ d between
superoutbursts).  This is more or less the full range of activity (and
brightness) displayed by dwarf novae.  Yet it occurs in three binaries
with orbital periods equal to within 2 minutes.  

Why this should be so remains a mystery.  Perhaps there are
cycles in the mass-transfer rate over timescales long compared 
to the historical record,
but short compared to the evolutionary timescale
\citep{kingmtcycle,wumtcycle}.  
Also, in the most common picture of evolution at short period,
the angular-momentum loss is thought to be dominated by 
gravitational radiation, leading to a convergence of evolution 
among different systems and a universal value for the minimum
period.  But if another, more idiosyncratic angular momentum loss
mechanism persists --- magnetic braking is an obvious candidate ---
the minimum period may not be as clearly defined, and the observed
diversity of stars near the minimum may be easier to understand.
Indeed, \citet{kingmtcycle} find that mass-transfer
cycles among short-period systems are damped out unless there
is some small consequential angular momentum loss (that is, an
angular momentum loss mechanism for which $\dot J$ increases
modestly with $\dot M$).  Another line of evidence also suggests an
extra channel for angular momentum loss.
\citet{patlate98} notes the apparent discrepancy between the large
number of short-period and `post-bounce' systems predicted by
theory, and the much smaller number of such systems which
have actually been observed; a second angular
momentum loss mechanism which destroys short-period systems
would help ameliorate that discrepancy, as well.

{\it Acknowledgments.} We thank the NSF for support
through grant AST 9987334, and the MDM staff for their 
support.  SV is a QEII fellow of the Australian Research Council.

\begin{deluxetable}{llcc}
\tablewidth{0pt}
\tablecolumns{6}
\tablecaption{Spectral Features}
\tablehead{
\colhead{Star} & \colhead{Feature} & \colhead{E.W.\tablenotemark{a}} 
& \colhead{Flux\tablenotemark{b}}  \\
 & & \colhead{(\AA )} & \colhead{$(10^{-16}$ erg cm$^{-2}$ s$^{1}$)}
}
\startdata
GW Lib & H$\beta$ (emission)  & 12  & 40  \\
       & FeII $\lambda$5169  & 1.3  & 6  \\
       & HeI $\lambda$5876  & 1.1  & 4 \\
       & Na D (emission) & 1.3 & 4  \\
       & H$\alpha$ (emission) & 52 & 120 \\
       & HeI $\lambda$6678  & 1.2  & 3 \\
V844 Her & H$\gamma$ & 40 & 160 \\ 
       & HeI $\lambda$4471  & 7  & 30  \\
       & H$\beta$  & 54  & 180  \\
       & HeI $\lambda$5015  & 5  & 16  \\
       & HeI $\lambda$5876  & 19  & 44  \\
       & H$\alpha$ & 98 & 200 \\
       & HeI $\lambda$6678  & 9.5  & 20  \\
DI UMa & H$\gamma$  & 20  & 60:   \\
       & H$\beta$  & 32  & 90  \\
       & HeI $\lambda$4921  & 2.5:  & 7:  \\
       & HeI $\lambda$5015  & 8.3  & 27  \\
       & HeI $\lambda$5876  & 10  & 22  \\
       & H$\alpha$ & 42 & 80 \\
\enddata
\tablenotetext{a}{Emission equivalent widths are counted as positive.}
\tablenotetext{b}{Absolute line fluxes are uncertain by a factor of about
2, but relative fluxes of strong lines
are estimated accurate to $\sim 10$ per cent.} 
\end{deluxetable}

\begin{deluxetable}{lrlrlrlr}
\tabletypesize{\scriptsize}
\tablewidth{0pt}
\tablecolumns{6}
\tablecaption{H$\alpha$ Radial Velocities}
\tablehead{
\colhead{HJD\tablenotemark{a}} & \colhead{V} & 
\colhead{HJD\tablenotemark{a}} & \colhead{V} & 
\colhead{HJD\tablenotemark{a}} & \colhead{V} & 
\colhead{HJD\tablenotemark{a}} & \colhead{V} \\
\colhead{} & \colhead{(km s$^{-1}$)} & 
\colhead{} & \colhead{(km s$^{-1}$)} &
\colhead{} & \colhead{(km s$^{-1}$)} &
\colhead{} & \colhead{(km s$^{-1}$)}
}
\startdata
\cutinhead{V844 Her:}
 $623.6793$ &  $  -87$   &  $623.8655$ &  $  137$   &  $624.7498$ &  $   -6$   &  $627.8050$ &  $   51$   \\ 
 $623.6834$ &  $  -80$   &  $623.8697$ &  $   61$   &  $624.7539$ &  $ -106$   &  $627.8091$ &  $   46$   \\ 
 $623.6876$ &  $ -141$   &  $623.8738$ &  $   69$   &  $624.7580$ &  $  -42$   &  $627.8131$ &  $   54$   \\ 
 $623.6917$ &  $  -41$   &  $623.8779$ &  $   10$   &  $624.7620$ &  $  -62$   &  $627.8191$ &  $  -78$   \\ 
 $623.6958$ &  $  -29$   &  $623.8821$ &  $  -52$   &  $624.7661$ &  $ -152$   &  $627.8232$ &  $ -136$   \\ 
 $623.7000$ &  $   84$   &  $623.8862$ &  $  -48$   &  $624.7701$ &  $ -107$   &  $627.8272$ &  $ -186$   \\ 
 $623.7041$ &  $ -112$   &  $623.8903$ &  $  -89$   &  $624.7742$ &  $ -149$   &  $627.8313$ &  $  -42$   \\ 
 $623.7082$ &  $   63$   &  $623.8945$ &  $ -149$   &  $624.9159$ &  $   -3$   &  $627.8353$ &  $  -47$   \\ 
 $623.7149$ &  $  -37$   &  $623.9052$ &  $  -84$   &  $624.9200$ &  $ -146$   &  $627.8394$ &  $  -31$   \\ 
 $623.7191$ &  $  -69$   &  $623.9093$ &  $   -8$   &  $624.9240$ &  $  -87$   &  $627.8434$ &  $    0$   \\ 
 $623.7232$ &  $ -132$   &  $623.9135$ &  $  -10$   &  $624.9281$ &  $ -169$   &  $631.8717$ &  $ -174$   \\ 
 $623.7273$ &  $ -213$   &  $623.9176$ &  $   65$   &  $624.9322$ &  $ -174$   &  $631.8758$ &  $ -102$   \\ 
 $623.7315$ &  $ -182$   &  $623.9217$ &  $   48$   &  $624.9362$ &  $ -175$   &  $631.8798$ &  $  -80$   \\ 
 $623.7356$ &  $  -84$   &  $623.9259$ &  $   17$   &  $624.9403$ &  $  -56$   &  $631.8839$ &  $    9$   \\ 
 $623.7397$ &  $  -23$   &  $623.9300$ &  $    5$   &  $624.9444$ &  $  -25$   &  $631.8879$ &  $  -29$   \\ 
 $623.7439$ &  $  -50$   &  $623.9341$ &  $  -61$   &  $624.9504$ &  $  -25$   &  $631.8920$ &  $   63$   \\ 
 $623.7509$ &  $  -68$   &  $623.9411$ &  $  -89$   &  $624.9544$ &  $  -70$   &  $631.8961$ &  $   -6$   \\ 
 $623.7550$ &  $   65$   &  $623.9452$ &  $ -168$   &  $624.9585$ &  $   80$   &  $631.9001$ &  $   56$   \\ 
 $623.7591$ &  $   48$   &  $623.9494$ &  $ -232$   &  $624.9625$ &  $   31$   &  $631.9068$ &  $   54$   \\ 
 $623.7633$ &  $  104$   &  $623.9535$ &  $ -196$   &  $624.9666$ &  $   95$   &  $631.9149$ &  $  -27$   \\ 
 $623.7674$ &  $  -60$   &  $623.9576$ &  $  -52$   &  $625.7567$ &  $ -144$   &  $631.9189$ &  $  -97$   \\ 
 $623.7716$ &  $ -111$   &  $623.9618$ &  $    5$   &  $625.7607$ &  $ -108$   &  $631.9230$ &  $ -130$   \\ 
 $623.7757$ &  $  -78$   &  $623.9659$ &  $   34$   &  $625.7648$ &  $  -90$   &  $631.9271$ &  $ -147$   \\ 
 $623.7798$ &  $ -152$   &  $624.6732$ &  $  -33$   &  $625.7689$ &  $  -33$   &  $631.9311$ &  $  -67$   \\ 
 $623.7919$ &  $ -116$   &  $624.6773$ &  $   49$   &  $625.7729$ &  $   16$   &  $631.9352$ &  $  -17$   \\ 
 $623.7961$ &  $  -55$   &  $624.6814$ &  $  -11$   &  $625.7770$ &  $  100$   &  $641.7186$ &  $  -85$   \\ 
 $623.8002$ &  $   15$   &  $624.6854$ &  $   41$   &  $625.7811$ &  $    9$   &  $641.7227$ &  $    9$   \\ 
 $623.8043$ &  $   25$   &  $624.6895$ &  $   22$   &  $625.7851$ &  $  -25$   &  $641.7267$ &  $   32$   \\ 
\tablebreak
 $623.8084$ &  $   44$   &  $624.6935$ &  $   97$   &  $625.7916$ &  $  -12$   &  $641.7308$ &  $   35$   \\ 
 $623.8126$ &  $   37$   &  $624.6981$ &  $  -33$   &  $625.7957$ &  $ -119$   &  $641.7348$ &  $   74$   \\ 
 $623.8167$ &  $   99$   &  $624.7021$ &  $  -73$   &  $625.7997$ &  $  -66$   &  $641.7389$ &  $    8$   \\ 
 $623.8208$ &  $   76$   &  $624.7100$ &  $  -54$   &  $625.8038$ &  $ -151$   &  $641.7457$ &  $   44$   \\ 
 $623.8289$ &  $  -75$   &  $624.7141$ &  $ -137$   &  $625.8078$ &  $  -84$   &  $641.7497$ &  $  -62$   \\ 
 $623.8330$ &  $ -139$   &  $624.7182$ &  $  -61$   &  $625.8119$ &  $ -157$   &  $641.7538$ &  $  -59$   \\ 
 $623.8371$ &  $ -157$   &  $624.7222$ &  $  -54$   &  $625.8160$ &  $  -73$   &  $641.7579$ &  $ -108$   \\ 
 $623.8413$ &  $ -136$   &  $624.7263$ &  $  -30$   &  $625.8200$ &  $  -44$   &  $641.7619$ &  $ -191$   \\ 
 $623.8454$ &  $ -108$   &  $624.7303$ &  $   -2$   &  $627.7888$ &  $   24$   &  $641.7660$ &  $ -100$   \\ 
 $623.8495$ &  $  -58$   &  $624.7344$ &  $  114$   &  $627.7928$ &  $   -7$   &  $641.7723$ &  $ -181$   \\ 
 $623.8537$ &  $   41$   &  $624.7385$ &  $   -7$   &  $627.7969$ &  $  160$   &  $641.7775$ &  $  -39$   \\ 
 $623.8578$ &  $  -31$   &  $624.7458$ &  $   29$   &  $627.8009$ &  $   76$   &  $641.7815$ &  $  156$   \\ 
\cutinhead{GW Lib:}
 $1333.7665$ &  $ -61$   &  $1339.8674$ &  $  -4$   &  $1340.8206$ &  $ -10$   &  $1341.7841$ &  $  37$   \\ 
 $1333.7705$ &  $ -52$   &  $1339.8714$ &  $  15$   &  $1340.8247$ &  $  -1$   &  $1341.7992$ &  $  -7$   \\ 
 $1333.7746$ &  $ -31$   &  $1340.6566$ &  $   7$   &  $1341.6564$ &  $ -45$   &  $1341.8033$ &  $  -7$   \\ 
 $1333.7787$ &  $ -28$   &  $1340.6607$ &  $ -10$   &  $1341.6605$ &  $ -57$   &  $1341.8073$ &  $ -50$   \\ 
 $1333.7828$ &  $  35$   &  $1340.6647$ &  $   5$   &  $1341.6645$ &  $ -35$   &  $1341.8114$ &  $ -33$   \\ 
 $1333.7869$ &  $   2$   &  $1340.6688$ &  $  20$   &  $1341.6686$ &  $ -36$   &  $1341.8154$ &  $ -52$   \\ 
 $1333.8434$ &  $  35$   &  $1340.6728$ &  $  -5$   &  $1341.6726$ &  $ -39$   &  $1341.8195$ &  $ -53$   \\ 
 $1333.8482$ &  $  40$   &  $1340.6769$ &  $   3$   &  $1341.6767$ &  $  14$   &  $1341.8235$ &  $ -52$   \\ 
 $1338.7317$ &  $ -62$   &  $1340.6899$ &  $ -28$   &  $1341.6807$ &  $  22$   &  $1341.8302$ &  $ -19$   \\ 
 $1338.7957$ &  $   9$   &  $1340.6940$ &  $ -28$   &  $1341.6953$ &  $   2$   &  $1341.8343$ &  $  14$   \\ 
 $1338.8479$ &  $   8$   &  $1340.6980$ &  $ -28$   &  $1341.6994$ &  $  -9$   &  $1341.8383$ &  $  19$   \\ 
 $1339.7590$ &  $  13$   &  $1340.7021$ &  $ -54$   &  $1341.7034$ &  $ -14$   &  $1341.8424$ &  $  42$   \\ 
 $1339.7666$ &  $   4$   &  $1340.7061$ &  $ -32$   &  $1341.7075$ &  $ -59$   &  $1341.8464$ &  $  -1$   \\ 
 $1339.7707$ &  $  34$   &  $1340.7102$ &  $ -16$   &  $1341.7115$ &  $ -41$   &  $1341.8505$ &  $  19$   \\ 
 $1339.7748$ &  $  12$   &  $1340.7143$ &  $   6$   &  $1341.7156$ &  $ -43$   &  $1341.8545$ &  $  -3$   \\ 
 $1339.7788$ &  $ -17$   &  $1340.7183$ &  $  -7$   &  $1341.7196$ &  $ -22$   &  $1341.8586$ &  $ -32$   \\ 
\tablebreak
 $1339.7829$ &  $ -19$   &  $1340.7963$ &  $ -67$   &  $1341.7598$ &  $ -80$   &  $1341.8627$ &  $ -34$   \\ 
 $1339.8452$ &  $ -70$   &  $1340.8004$ &  $ -76$   &  $1341.7639$ &  $ -40$   &  $1341.8667$ &  $ -60$   \\ 
 $1339.8512$ &  $ -53$   &  $1340.8044$ &  $ -76$   &  $1341.7679$ &  $ -82$   &  $1341.8708$ &  $ -28$   \\ 
 $1339.8552$ &  $ -43$   &  $1340.8085$ &  $ -58$   &  $1341.7720$ &  $ -39$   &  $1341.8748$ &  $ -57$   \\ 
 $1339.8593$ &  $ -21$   &  $1340.8125$ &  $ -46$   &  $1341.7760$ &  $ -24$   &  $1341.8789$ &  $ -27$   \\ 
 $1339.8633$ &  $ -22$   &  $1340.8166$ &  $ -18$   &  $1341.7801$ &  $   1$   &  $1341.8829$ &  $ -37$   \\ 
\enddata
\tablenotetext{a}{Heliocentric JD of mid-integration minus 2450000.}
\end{deluxetable}

\begin{deluxetable}{lrrrrc}
\footnotesize
\tablewidth{0pt}
\tablecaption{Fits to Radial Velocities}
\tablehead{
\colhead{Star} & \colhead{$T_0$\tablenotemark{a}} & \colhead{$P$} &
\colhead{$K$} & \colhead{$\gamma$} & \colhead{$\sigma$}  \\
\colhead{} & \colhead{} &\colhead{(d)} & \colhead{(km s$^{-1}$)} &
\colhead{(km s$^{-1}$)} & \colhead{(km s$^{-1}$)}
}
\startdata
GW Lib & $1340.6580 \pm 0.0007$ & $0.05332 \pm 0.00002$  & $38 \pm 3$ 
& $ -17 \pm 2$ & 16 \\ 
V844 Her & $631.6119 \pm 0.0010$ & $0.054643 \pm 0.000007$ & $97 \pm 11$ 
& $-38 \pm 8$ & 47 \\
\enddata
\tablenotetext{a}{Apparent emission-line inferior conjunction, HJD $- 2450000$.}
\end{deluxetable}

\begin{deluxetable}{lrrcccrr}
\footnotesize
\tablewidth{0pt}
\tablecolumns{8}
\tablecaption{Photometry Log (V844 Her)}
\tablehead{
\colhead{UT Date} & 
\colhead{Start\tablenotemark{a}} &
\colhead{Duration} & 
\colhead{$N_{\rm points}$} &
\colhead{Tel.\tablenotemark{b}} &
\colhead{$\left<V\right>$} & 
\colhead{$T_{\rm max}$\tablenotemark{c}} & 
\colhead{Amplitude} \\
\colhead{(1997)} & 
\colhead{} &
\colhead{(h)} & 
\colhead{} &
\colhead{} &
\colhead{(mag)} &
\colhead{} &
\colhead{(mag)} \\ 
}
\startdata
May 24  & 592.4162 &  4.10 & 243 & 1      &13.09  & 592.4604  & 0.18 \\
May 25  & 593.4381 &  2.55 &  75 & 2      &13.17  & 593.4637  & 0.14 \\
May 26  & 594.3971 &  4.97 & 250 & 1      &13.27  & 594.4130  & 0.11 \\
May 28  & 596.5536 & 10.43 & 773 & 3,4    &13.54  & 596.5984  & 0.067 \\
May 29  & 597.6080 &  5.90 & 375 & 3      &13.62  & 597.6625  & 0.036 \\
May 30  & 598.7468 &  5.61 & 222 & 4      &       & 598.7860  & 0.062 \\
May 31  & 599.4120 & 13.72 & 870 & 1,2,4,5&13.70  & 599.4570  & 0.12 \\
Jun 1  & 600.4180 & 12.97 & 716 & 2,3,4  &13.72  & 600.4600  & 0.09 \\
Jun 2 & 601.7096 &  6.68 & 348 & 4      &13.90  & 601.7446  & 0.064 \\
Jun 3 & 602.6403 &  7.68 & 400 & 4      &14.04  & 602.6921  & 0.058 \\
Jun 4 & 603.6619 &  7.12 & 213 & 5   &15.2-15.7& 603.6970  & 0.062 \\
Jun 5 & 604.6566 &  7.44 & 471 & 4,5    &17.00  &           & $<0.05$ \\
Jun 10 & 609.6602 &  6.92 & 189 & 4      &17.46  & 609.6672  & 0.18 \\
Jun 11 & 610.6594 &  7.71 & 214 & 4      &17.51  & 610.6734  & 0.23 \\
Jun 14 & 613.6465 &  7.13 & 190 & 4      &17.72  & 613.6678  & 0.15 \\
Jun 15 & 614.6075 &  8.28 & 376 & 3,4    &17.68  & 614.6769  & 0.22 \\
Jun 16 & 615.6546 &  7.10 & 182 & 4      &17.77  & 615.6703  & 0.16 \\
Jun 17 & 616.6607 &  7.17 & 200 & 4      &18.00  & 616.6729  & 0.16 \\
Jun 20 & 619.5792 &  4.97 & 266 & 3      &18.10  & 619.6075  & 0.16 \\
Jun 28 & 627.5614 &  5.12 & 307 & 3      &17.41  &           & $<0.04$ \\
Jun 30 & 629.5679 &  2.12 & 114 & 3      &17.28  & & \\
\enddata
\tablenotetext{a}{Heliocentric JD of start, minus 2450000.}
\tablenotetext{b}{Telescope codes: 
1 = CBA-Belgium (25 cm); 2 = CBA-Denmark (25 cm);
3 = CBA-Maryland (66 cm); 4 = CBA-Tucson (35 cm);
5 = CBA-Braeside (41 cm).}
\tablenotetext{c}{Heliocentric JD of superhump maximum, minus 2450000.}
\end{deluxetable}

\begin{deluxetable}{lllll}
\footnotesize
\tablewidth{0pt}
\tablecolumns{5}
\tablecaption{Short-Period Dwarf Novae}
\tablehead{
\colhead{Star} &
\colhead{$P_{\rm orb}$} &
\colhead{$P_{\rm sh}$\tablenotemark{a}} &
\colhead{$\epsilon$} &
\colhead{Source} \\
\colhead{} &
\colhead{(d)} &
\colhead{(d)} &
\colhead{} & 
\colhead{}  
}
\startdata
V485 Cen &  0.040995(1) &  0.04216(2) & 0.0284(5) &  \citet{aug96} \\
       &&&&                                         \citet{olech97} \\
RX2329+06 & 0.0445671(2) & 0.0462(1) & 0.037(3) & \citet{thor02} \\
       &&&&                                       \citet{skill02} \\
GW Lib   &  0.05332(2)       &  \nodata   &  \nodata    &    This work \\ 
RX2353$-$38 & 0.0543234(?)  &  \nodata  &  \nodata      &    \citet{aw97}; \\
	& & & & \citet{ab97}  \\
DI UMa  &  0.054564(2) &  0.05529(3) & 0.0133(5)  & 
\citet{kato96}; \\ 
 	& & & & \citet{fr99} \\
V844 Her & 0.054643(7) &  0.05597(5)&  0.0243(9) &  This work  \\      
V592 Her &  \nodata   &   0.05653(13)  &  \nodata     &   \citet{dm98}\\ 
\enddata 
\tablenotetext{a}{
Superhump periods are slightly unstable.  We adopt 
the mean value, or in the case of sparse data,
an estimate 4 d after superhump onset.  This reproduced the mean value for
stars where both quantities could be observed.  The estimated error in 
$P_{\rm sh}$ the error in estimating this quantity, not the full range
of variation in $P_{\rm sh}$ (which is considerably larger).
}
\end{deluxetable}

\clearpage

\begin{figure}
\plotone{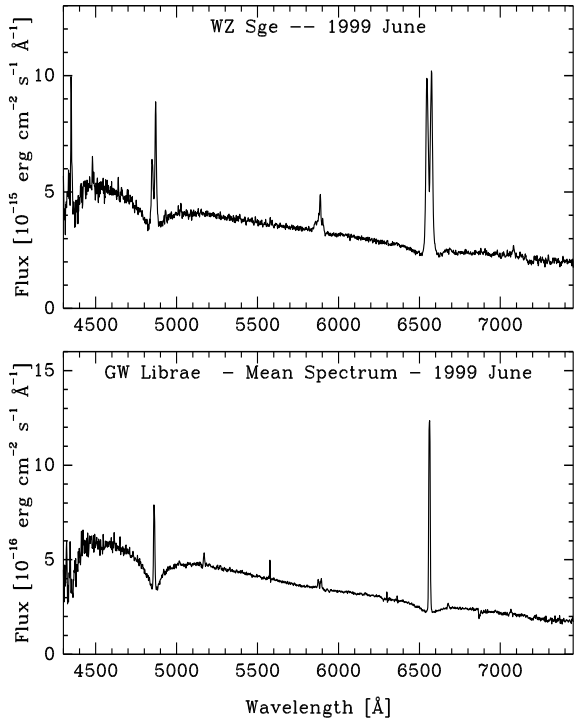}
\caption{{\it Lower panel:} Mean spectrum of GW Lib. 
{\it Upper panel:} Spectrum of WZ Sge obtained for comparison.
Note the similar broad Balmer absorption profiles from the white
dwarf, and the markedly sharper emission lines in GW Lib.
}
\end{figure}

\begin{figure}
\plotone{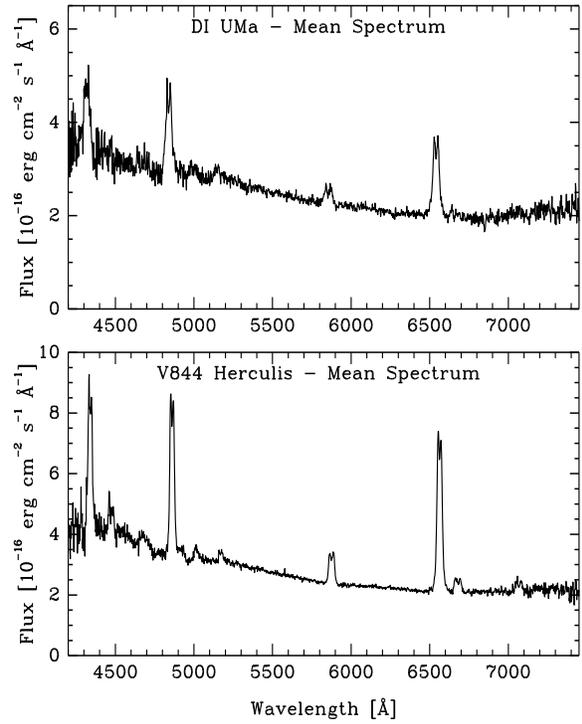}
\caption{{\it Lower panel:} Mean spectrum of V844 Her.
{\it Upper panel:} Mean spectrum of DI UMa, based on 
19 6-minute exposures obtained 1997 Dec 17.510 - 17.556 UT.
}
\end{figure}

\clearpage

\begin{figure}
\plotone{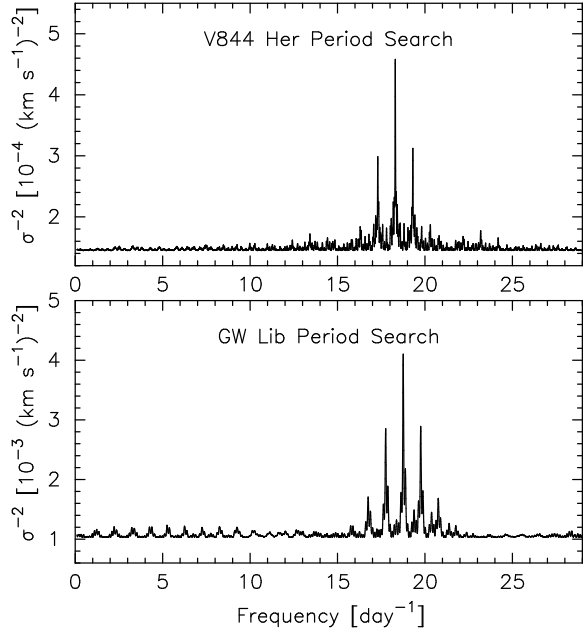}
\caption{{\it Lower panel:} Period search `residualgram' of the
GW Lib H$\alpha$ emission velocities. {\it Upper panel:} Period
search of the V844 Her H$\alpha$ emission velocities.
}
\end{figure}

\begin{figure}
\plotone{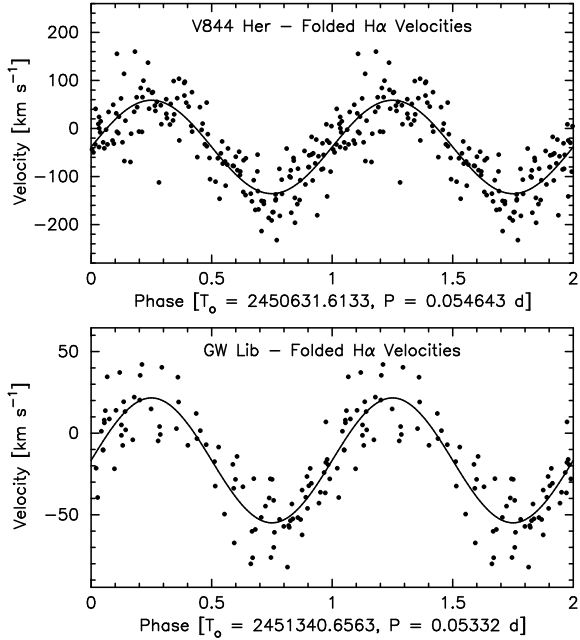}

\caption{H$\alpha$ velocities of 
GW Lib and V844 Her folded on the adopted periods,
with best-fitting sinusoids superposed.  All data are
plotted twice to maintain continuity.
}
\end{figure}

\clearpage
\begin{figure}   
\plotone{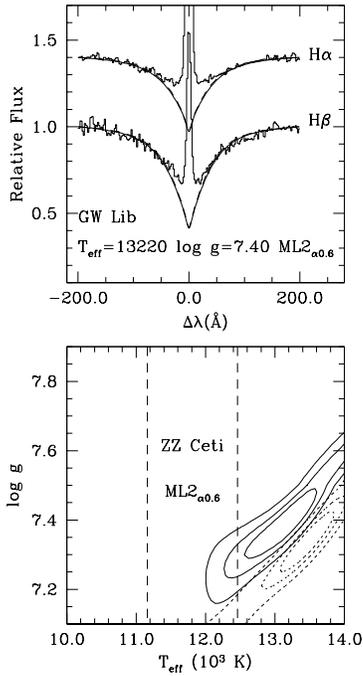}
\caption{{\it Upper:} H $\alpha$ and H$\beta$ profiles, with 
the best-fitting model atmosphere calculations superposed.  {\it Lower:}
contours of equal $\chi^2$ on the $\log g$-$T_{\rm eff}$ plane.  
For the solid contours, a $\pm 50$ \AA\ interval around the line
cores was excluded; the best fit is at 
($T_{\rm eff}, \log g) = (13220\  {\rm kelvin}, 7.4)$.
For the dashed contours, $\pm 40$ and $\pm 30$ \AA\ intervals
were excluded at H$\alpha$ and H$\beta$ respectively, and the 
best fit is at ($13460\  {\rm kelvin}, 7.34)$. 
}
\end{figure}

\begin{figure}   
\plotone{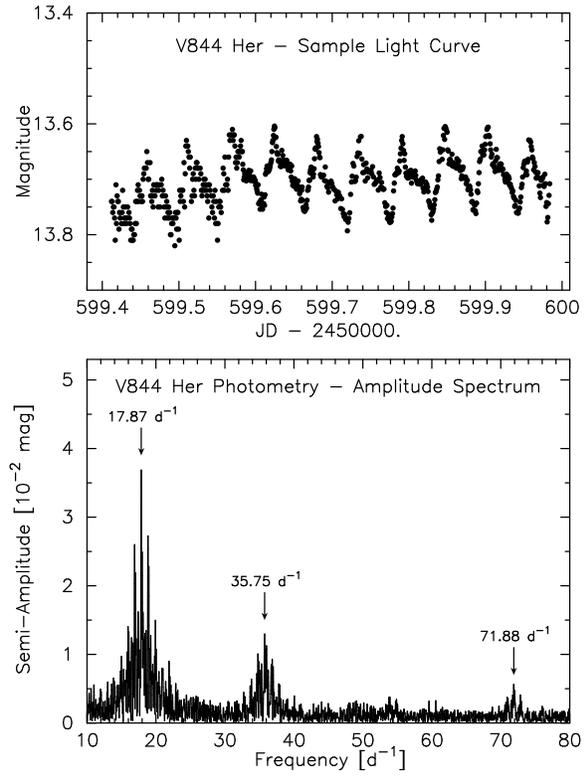}
\caption{{\it Upper panel:} Superhumps observed in V844 Her 
on a single long night of photometric observation.  
{\it Lower panel:} Amplitude spectrum of 10 days of outburst
photometry.  The frequencies noted in the text are indicated
with arrows.}
\end{figure}

\end{document}